\documentclass[aps,prl,twocolumn,showkeys,superscriptaddress]{revtex4-1}
\usepackage{amsmath,amssymb,bm,graphicx,color}

\begin{document}

\title{Optical Phonon Lasing in Semiconductor Double Quantum Dots}

\author{Rin Okuyama}
\email{rokuyama@rk.phys.keio.ac.jp}
\author{Mikio Eto}
\affiliation{Faculty of Science and Technology,
    Keio University, Yokohama 223-8522, Japan}
\author{Tobias Brandes}
\affiliation{Institut f\"ur Theoretische Physik,
    Technische Universit\"at Berlin, D-10623 Berlin, Germany}
\date{\today}

\begin{abstract}
We propose optical phonon lasing for a double quantum dot (DQD)
fabricated in a semiconductor substrate. We show that
the DQD is weakly coupled
to only two LO phonon modes that act as a natural cavity.
The lasing occurs for pumping the DQD via electronic tunneling
at rates much higher than the phonon decay rate, whereas an
antibunching of phonon emission is observed in the opposite regime
of slow tunneling. Both effects disappear with an effective thermalization
induced by the Franck-Condon effect in a DQD fabricated in a carbon
nanotube with a strong electron-phonon coupling.
\end{abstract}


\maketitle

Electrically tunable two-level systems are ideal candidates to study
the interaction between fermions and bosons under nonequilibrium conditions.
Single-qubit lasers or nanomechanical resonators
    \cite{Astafiev2007,Brandes2003,Blanter2004,Rodrigues2007,
    Huebener2007,Andre2009,Gartner2011}
are examples where  concepts from  quantum optics,
such as the microlaser
    \cite{McKeever2003},
have been successfully transferred to and combined with artificial
solid-state architectures.
Semiconductor double quantum dots (DQDs) play a similar role as model
systems with the coupling between electrons and the surrounding substrate
leading to, e.g., tunable spontaneous phonon emission and Dicke-type
interference effects
    \cite{Fujisawa1998,Brandes1999,Roulleau2011}.

In this Letter, we propose optical phonon lasing in a DQD
without the requirement of an additional cavity or resonator.
We start from the observation that a DQD effectively couples to only
two LO phonon modes that work as a natural cavity.
The pumping to the upper level is realized by an electric current
through the DQD under a finite bias.
The amplified LO phonons occasionally escape from this cavity
by decaying into the so-called ``daughter phonons''
    \cite{Vallee1994}
that can be observed externally. The phonon lasing is possible
when the pumping rate is much higher than the phonon decay rate
$\Gamma_{\rm ph}$.

We also observe the phonon antibunching in the same system when the
pumping rate is lower than $\Gamma_{\rm ph}$.
We emphasize that the phonon statistics can be changed by
electrically tuning the tunnel coupling between DQDs and leads
    \cite{Jin2012}.
Note that LO-phonon-assisted transport through a DQD was
theoretically studied
by Gnodtke {\it et al}.\
    \cite{Gnodtke2006}
and has recently been observed by Amaha and Ono
    \cite{Amaha2012}.
We also note that phonon lasing by optical pumping
was suggested in single quantum dots
    \cite{Kabuss2012}.

Both the phonon lasing and antibunching are 
weak coupling effects that are spoilt
by phonon thermalization via the Franck-Condon effect
    \cite{Koch2005,Sapmaz2006,Leturcq2009}
for a strong electron-phonon coupling.
In electric transport, the number of electrons
in the DQD fluctuates, which is accompanied by lattice distortions
and thus the creation of bunched phonons.
We show that this effect is negligible in DQDs
fabricated on GaAs substrates (weak coupling case) but surpasses
the lasing and antibunching in DQDs on carbon nanotubes (CNTs;
strong coupling case)
    \cite{com:coupling_strength}.

\begin{figure}[t]
\begin{center}
    \includegraphics[width=8.5cm]{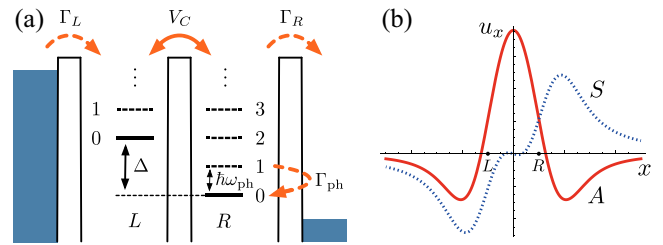}
    \caption{(Color online).
    (a) Model for a double quantum dot (DQD) coupled to LO
    phonons. A large bias is applied between external leads.
    The spacing $\Delta$ between
    the energy levels in dots $L$ and $R$ is electrically tunable.
    When $\Delta$ matches an integer ($\nu$) multiple of the phonon
    energy $\hbar \omega_{\rm ph}$, the electronic state $|L \rangle$ with
    $n$ phonons is coherently coupled to $|R \rangle$ with
    $(n+\nu)$ phonons.
    (b) Phonon mode functions ${\bm u}({\bm r})$
    along a line through the centers of quantum dots
    located at
    $x=\pm \mathcal{R}$, when the wavefunctions $|L \rangle$
    and $|R \rangle$ are spherical with radius $\mathcal{R}$.
    The $x$-component of ${\bm u}(x,0,0)$ is shown for
    $S(A)$-phonons that couple (anti-)symmetrically to the DQD.
    Note that $u_x$ is an odd (even) function of $x$ for
    $S(A)$-phonons since the induced charge is proportional to
    $\bm{\nabla} \cdot \bm{u}$.
    \label{fig:model}}
\end{center}
\end{figure}

Figure \ref{fig:model}(a) depicts our model of a DQD embedded
in a substrate, in which two single-level quantum dots, $L$ and $R$, are
connected by tunnel coupling $V_C$.
The level spacing $\Delta$ between the dots is assumed to be tunable,
and the total number of electrons
in the DQD is restricted to one or zero
by the Coulomb blockade. The electron couples to
LO phonons of energy $\hbar \omega_{\rm ph}$ in the substrate
via the Fr\"ohlich interaction. Our system Hamiltonian is
$
    \mathcal{H} = \mathcal{H}_e + \mathcal{H}_{\rm ph}
        + \mathcal{H}_{\rm ep}
$,
\begin{align}
    \mathcal{H}_e &= \frac{\Delta}{2} (n_L - n_R)
        + V_C (d_L^\dagger d_R + d_R^\dagger d_L), \\
    \mathcal{H}_{\rm ph} &= \hbar \omega_{\rm ph} \sum_{\bm{q}} N_{\bm{q}},
        \\
    \mathcal{H}_{\rm ep} &=  \sum_{\alpha = L, R} \sum_{\bm{q}}
       M_{\alpha, \bm{q}} (a_{\bm{q}} + a_{-\bm{q}}^\dagger) n_\alpha,
\end{align}
using creation (annihilation) operators
$d_{\alpha}^{\dagger}$ ($d_{\alpha}$) for an electron in
dot $\alpha$ and $a_{\bm{q}}^\dagger$ ($a_{\bm{q}}$) for
a phonon with wave vector $\bm{q}$.
$n_{\alpha} = d_{\alpha}^\dagger d_{\alpha}$ and
$N_{\bm{q}} = a_{\bm{q}}^\dagger a_{\bm{q}}$ are the
number operators.
The spin index is omitted for electrons.
The coupling constant is given by
$
    M_{\alpha, \bm{q}} =
    \sqrt{\frac{e^2 \hbar \omega_{\rm ph}}{2V}(\frac{1}{\epsilon(\infty)}
    - \frac{1}{\epsilon(0)})}
    \frac{1}{q} \langle \alpha | e^{i \bm{q} \cdot \bm{r}} | \alpha \rangle
$,
where $| \alpha \rangle$ is the electron wavefunction in dot $\alpha$,
$\epsilon(\infty)$ [$\epsilon(0)$] is the dielectric constant
at a high [low] frequency, and $V$ is the substrate volume.
The LO phonons only around the $\Gamma$ point ($|\bm{q}| \lesssim
1/\mathcal{R}$ with dot radius $\mathcal{R}$)
are coupled to the DQD
because of an oscillating factor in $M_{\alpha, \bm{q}}$.
This fact justifies the dispersionless phonons in $\mathcal{H}_{\rm ph}$.
We assume equivalent quantum dots $L$ and $R$, whence
$M_{R, \bm{q}}=M_{L, \bm{q}} e^{i \bm{q} \cdot \bm{r}_{LR}}$ with
$\bm{r}_{LR}$ being a vector joining their centers.

In $\mathcal{H}_{\rm ep}$,
an electron in dot $\alpha$ couples to a single mode of a phonon described by
\begin{align}
    a_\alpha = \frac{\sum_{\bm{q}} M_{\alpha, \bm{q}} a_{\bm{q}}}
        {(\sum_{\bm{q}} |M_{\alpha, \bm{q}}|^2)^{1/2}}.
   \label{eq:a_La_R}
\end{align}
We perform a unitary transformation for phonons from $a_{\bm{q}}$
to collective phonon coordinates,
\begin{align}
    a_{S} = \frac{a_{L} + a_{R}}{\sqrt{2 (1 + \mathcal{S})}},
\quad
    a_{A} = \frac{a_{L} - a_{R}}{\sqrt{2 (1 - \mathcal{S})}},
\end{align}
and other modes orthogonal to $a_S$ and $a_A$, where $\mathcal{S}$
is the overlap integral between $a_L$ and $a_R$ phonons
in Eq.\ (\ref{eq:a_La_R}).
Disregarding the modes decoupled from the DQD,
we obtain the effective Hamiltonian
\begin{align}
    H=\mathcal{H}_e + &
      \hbar \omega_{\rm ph}
        \left[ N_S + \lambda_S (a_S+ a_S^\dagger)
                            (n_L + n_R) \right]
      \nonumber \\
      + & \hbar \omega_{\rm ph}
        \left[ N_A + \lambda_A (a_A+ a_A^\dagger)
                            (n_L - n_R) \right],
        \label{eq:H_eff}
\end{align}
with $N_S = a_{S}^\dagger a_{S}$, $N_A = a_{A}^\dagger a_{A}$,
and dimensionless coupling constants 
$
\lambda_{S/A} = ( \sum_{\bm{q}} |
M_{L, \bm{q}} \pm M_{R, \bm{q}}|^2)^{1/2}/(2 \hbar \omega_{\rm ph})
$.

The mode functions for $S$- and $A$-phonons are depicted in Fig.\ 
\ref{fig:model}(b)
along a line through the centers of the quantum dots.
Since the phonons are dispersionless, they
do not diffuse and act as a cavity including the DQD
    \cite{com:flat_dispersion}.
$A$-phonons play a crucial role in phonon-assisted tunneling and phonon
lasing, as discussed below, whereas $S$-phonons do not since they
couple to the {\em total} number of electrons in the DQD, $n_L + n_R$.
Both phonons are relevant to the Franck-Condon effect.

Our Hamiltonian $H$ in Eq.\ (\ref{eq:H_eff})
is applicable to  DQDs fabricated on a semiconductor substrate, where
$\hbar \omega_{\rm ph} = 36~\text{meV}$
and $\lambda_{S,A} = 0.1-0.01$ for
$\mathcal{R} = 10-100~\text{nm}$ in GaAs
    \cite{Tasai2003}.
It also describes a DQD in a suspended
CNT when an electron couples to a vibron, which is a longitudinal
stretching mode with $\hbar \omega_{\rm ph} \sim 1~\text{meV}$,
$\lambda_A \gtrsim 1$,
and $\lambda_S = 0$ in experimental situations
    \cite{Sapmaz2006, Leturcq2009}.

The DQD is connected to external leads in series, which enables
electronic pumping.
Under a large bias, an electron tunnels into dot $L$ from the
left lead with
tunneling rate $\Gamma_{L}$ and tunnels out from dot $R$ to the
right lead with $\Gamma_{R}$
    \cite{com:large_bias}.
We also introduce the phonon decay rate $\Gamma_{\rm ph}$ to
take into account the natural decay of LO phonons due to lattice
anharmonicity
    \cite{Vallee1994}.
We describe the dynamics of the DQD-phonon density matrix $\rho$
using the Markovian master equation
\begin{align}
    \dot \rho &= - \frac{i}{\hbar} [H, \rho]
        + \mathcal{L}_e \rho + \mathcal{L}_{\rm ph} \rho,
        \label{eq:master}
\end{align}
where $
    \mathcal{L}_e \rho = ( \Gamma_L \mathcal{D} [d_L^\dagger]
    + \Gamma_R \mathcal{D} [d_R] ) \rho
$ and
$
    \mathcal{L}_{\rm ph} \rho = \Gamma_{\rm ph}
    (\mathcal{D} [a_S] + \mathcal{D} [a_A]) \rho
$,
with $\mathcal{D}[A] \rho = A \rho A^\dagger
- \frac{1}{2} \{ \rho, A^\dagger A \}$ being a Lindblad dissipator.
Here, we assumed that the temperature of the substrate, $k_B T$,
is much lower than $\hbar \omega_{\rm ph}$.

In the following, 
we apply the Born-Markov-Secular approximation to Eq.\ (\ref{eq:master})
by diagonalizing the Hamiltonian $H$ and setting up the corresponding
rate equation in the energy eigenbasis,
\begin{align}
    \dot{P_i} &= \sum_{j} L_{ij} P_j,
    \label{eq:rate}
\end{align}
for the probabilities $P_i$ to find the system
in an eigenstate $|i\rangle$, with  
$
    L_{ij} = \langle i| [
        (\mathcal{L}_e + \mathcal{L}_{\rm ph}) |j \rangle \langle j|
    ] |i \rangle.
$
The solution of Eq.\ (\ref{eq:rate}) with $\dot{P_i}=0$
determines the steady state properties.

\begin{figure}
\begin{center}
    \includegraphics[width=7.5cm]{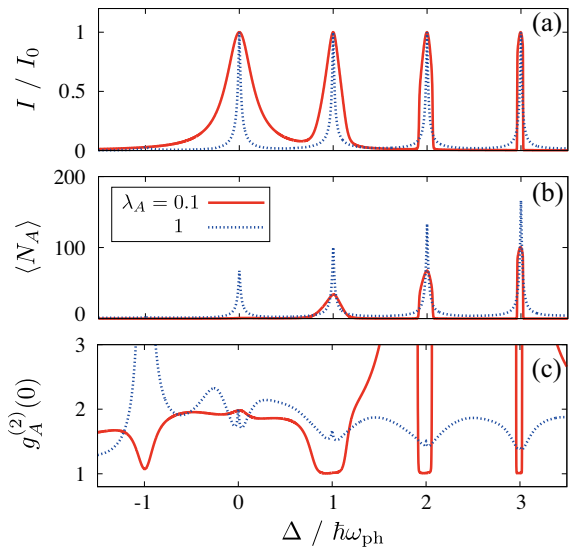}
    \caption{(Color online).
    (a) Electric current through the DQD,
    (b) $A$-phonon number $\langle N_A \rangle$,
    and (c) its autocorrelation function
    $g^{(2)}_A (0)$ as a function of level spacing $\Delta$ in the
    DQD. The dimensionless electron-phonon couplings are
    $\lambda_A = 0.1$ (solid lines) or $1$ (dotted lines),
    and $\lambda_S = 0$.
    $I_0 = e \Gamma_R / (2 + \alpha)$ is the current at $\Delta=0$
    in the absence of electron-phonon coupling.
    $\Gamma_L = \Gamma_R = 100~\Gamma_{\rm ph}$
    and $V_C = 0.1~\hbar \omega_{\rm ph}$.
    \label{fig:Delta}}
\end{center}
\end{figure}

First, we discuss our numerical results in the case of
$\Gamma_{L, R} \gg \Gamma_{\rm ph}$. We consider $A$-phonons
and disregard $S$-phonons ($\lambda_S=0$).
Figure \ref{fig:Delta}(a) shows the current $I$ through the DQD as
a function of level spacing $\Delta$, with $\lambda_A=0.1$ and $1$.
Beside the main peak at $\Delta = 0$, we observe
subpeaks at $\Delta \simeq \nu \hbar \omega_{\rm ph}$
($\nu = 1, 2, 3, \ldots$) due to the phonon-assisted
tunneling. At the $\nu$th subpeak, electron transport through
the DQD is accompanied by the emission of $\nu$ phonons.
As a result, the phonon number is markedly increased at the
subpeaks, as shown in Fig.\ \ref{fig:Delta}(b), in both cases of
$\lambda_A=0.1$ and $1$. However, the physics is very different for
the two cases, as we will show below.

For $\lambda_A=0.1$ and $\Delta \simeq \nu \hbar \omega_{\rm ph}$,
the electronic state
$| L \rangle$ with $n$ phonons is coherently coupled to
$| R \rangle$ with $(n+\nu)$ phonons
\cite{Hameau1999}, similarly to  a microlaser two-level system in
a photon cavity, if the lattice distortion can be neglected. 
To examine the amplification of $A$-phonons,
we calculate the phonon autocorrelation function
\begin{align}
    g^{(2)}_A (\tau) = \langle :N_A (0) N_A (\tau): \rangle
    / \langle N_A \rangle^2,
    \label{eq:g2func}
\end{align}
which is the probability of phonon emission at
time $\tau$ on the condition that a phonon is emitted at time $0$
\cite{Scully, Emary2012}. $g^{(2)}_A (0)=1$ indicates a
{\em Poissonian} distribution of phonons,
which is a criterion of phonon lasing.
We thus find phonon lasing at the current subpeaks
in Fig.\ \ref{fig:Delta}(c).
We mention that these are not changed in the
presence of finite coupling $\lambda_S$ to $S$-phonons.

When $\lambda_A=1$, the strength of the electron-phonon
interaction is comparable to the phonon energy. In this case, the
lattice distortion by the Franck-Condon effect severely disturbs
the above-mentioned coherent coupling between an electron and phonons
in the DQD and, as a result, suppresses the phonon lasing.
Indeed, $g^{(2)}_A (0)>1$ at the current subpeaks, indicating
the phonon bunching.

\begin{figure}[t]
\begin{center}
    \includegraphics[width=7.5cm]{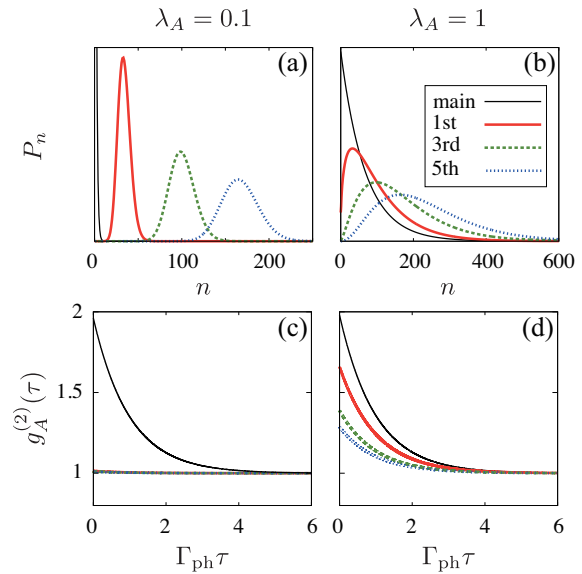}
    \caption{(Color online).
    (a), (b) Number distribution
    of $A$-phonons and (c), (d) its autocorrelation function
    $g^{(2)}_A (\tau)$ at the current main peak and
    subpeaks ($\nu=1,3,5$).
    The dimensionless electron-phonon couplings are
    $\lambda_A = 0.1$ [$1$] in panels (a) and (c)
    [(b) and (d)] and $\lambda_S = 0$.
    Note that three lines for the current subpeaks
    are almost overlapped in panel (c).
    $\Gamma_L = \Gamma_R = 100~\Gamma_{\rm ph}$ and
    $V_C = 0.1~\hbar \omega_{\rm ph}$.
    \label{fig:distribution}}
\end{center}
\end{figure}

To compare the two situations in detail,
we show the number distribution of $A$-phonons in Figs.\
\ref{fig:distribution}(a) and \ref{fig:distribution}(b)
at the current main peak and subpeaks. In the case of $\lambda_A = 0.1$,
a Poisson-like distribution emerges at the subpeaks,
whereas a Bose distribution with effective temperature $T^*$
is seen at the main peak.
$T^*$ is determined from the number of phonons in the stationary state as
$1/[e^{\hbar \omega_{\rm ph} / (k_B T^*)} - 1] = \langle N_A \rangle$.
When $\lambda_A = 1$, on the other hand,
the distribution shows an intermediate shape between
Poissonian and Bose distributions at the subpeaks
and a Bose distribution at the main peak.
In Figs.\ \ref{fig:distribution}(c) and \ref{fig:distribution}(d),
we plot the autocorrelation function $g^{(2)}_A (\tau)$ of
$A$-phonons as a function of $\tau$.
In the case of $\lambda_A = 0.1$, $g^{(2)}_A (\tau) \simeq 1$,
regardless of the time delay $\tau$,
supports the phonon lasing at the current subpeaks.
At the main peak,
$g^{(2)}_A (\tau) \simeq 1 + e^{- \Gamma_{\rm ph} \tau}$, which is
a character of thermal phonons with effective temperature $T^*$.
When $\lambda_A = 1$, we find an intermediate behavior,
$g^{(2)}_A (\tau) \simeq 1 + \delta_\nu e^{- \Gamma_{\rm ph} \tau}$
($0 < \delta_\nu < 1$), at the $\nu$th subpeak.
This indicates that the phonons are partly thermalized
by the Franck-Condon effect. For larger $\nu$ values, the distribution
is closer to the Poissonian with smaller $\delta_\nu$.

To elucidate the competition between the phonon lasing and
thermalization by the Franck-Condon effect, we analyze the
rate equation in Eq.\ (\ref{eq:rate}), focusing on the current peaks.
We introduce  polaron states
$|L(R), n \rangle$ for an electron in dot $L$ $(R)$ and $n$ phonons
with lattice distortion:
\begin{equation}
    |L(R), n \rangle = |L(R) \rangle
        \otimes \mathcal{T}^{(\dagger)} |n \rangle_A,
        \quad \mathcal{T} = e^{- \lambda_A (a_A^\dagger - a_A)},
\end{equation}
where $\mathcal{T}$
and $\mathcal{T}^\dagger$
describe the shift of the equilibrium position of the lattice
when an electron stays in dots $L$ and $R$, respectively.
Note that the lattice distortion produces $\lambda_A^2$ extra
phonons: $\langle \alpha, n| N_A |\alpha, n \rangle = n + \lambda_A^2$.
When $\Delta \simeq \nu \hbar \omega_{\rm ph}$,
the eigenstates of Hamiltonian $H$ are given by
the zero-electron states $|0, n \rangle = |0 \rangle \otimes |n \rangle_A$
and bonding and antibonding states
between the polarons
\begin{eqnarray}
    |\pm, n \rangle = \frac{1}{\sqrt2}
        (|L, n \rangle \pm |R, n + \nu \rangle)
\end{eqnarray}
$(n = 0, 1, 2, \ldots)$ and the polarons localized in dot $R$,
$|R, n \rangle$ ($n = 0, 1, 2, \ldots, \nu - 1$),
in a good approximation,
provided that $V_C \ll \hbar \omega_{\rm ph}$.
The rate equations for these states are
\begin{align}
    \dot{P}_{0, n} = &-\Gamma_L P_{0, n}
        + \sum_{m = 0}^{\infty} \frac{\Gamma_R}{2}
        |_A \langle n | \mathcal{T}^\dagger | m + \nu \rangle_A |^2
        P_{\text{mol}, m}
        \nonumber \\
    &+ \sum_{m = 0}^{\nu - 1} \Gamma_R
        |_A \langle n | \mathcal{T}^\dagger | m \rangle_A |^2 P_{R, m}
        \nonumber \\
    &+ \Gamma_{\rm ph} \left[ (n + 1) P_{0, n + 1} - n P_{0, n} \right],
        \\
    \dot{P}_{\text{mol}, n} = &- \frac{\Gamma_R}{2} P_{\text{mol}, n}
        + \sum_{m = 0}^{\infty} \Gamma_L
        |_A \langle n | \mathcal{T} | m \rangle_A |^2 P_{0, m}
        \nonumber \\
    &+ \Gamma_{\rm ph} \left[
        \left( n + 1 + \frac{\nu}{2} \right) P_{\text{mol}, n + 1}
        - \left( n + \frac{\nu}{2} \right)
        P_{\text{mol}, n} \right],
        \label{eq:rate_mol}
\end{align}
where $P_{\text{mol}, n} = P_{+, n} + P_{-, n}$
($n = 0, 1, 2, \ldots$) and
\begin{align}
    \dot{P}_{R, n} &= - \Gamma_R P_{R, n}
        + \Gamma_{\rm ph} \bigl[ (n + 1) P_{R, n + 1} - n P_{R, n} \bigr],
        \label{eq:rate_R}
\end{align}
with $P_{R, \nu} = P_{\text{mol}, 0}/2$ ($n = 0, 1, 2, \ldots, \nu - 1$).
These equations yield the current $I$ and the
electron number in the DQD
$
    \langle n_e \rangle = \langle n_L + n_R \rangle
$
in terms of the number of polarons localized in dot $R$,
$
    \langle \tilde n_R \rangle = \sum_{n = 0}^{\nu - 1} P_{R, n}
$ as 
\begin{align}
    I = e \Gamma_R \frac{ 1 + \langle \tilde n_R \rangle }{2 + \alpha},
        \quad
    \langle n_e \rangle= \frac{2 - \alpha \langle \tilde n_R \rangle}
                            {2 + \alpha}
        \label{eq:I}
\end{align}
with $\alpha = \Gamma_R / \Gamma_L$.
The number of $A$-phonons is given by
\begin{align}
    \langle N_A \rangle = (\nu + 2 \lambda_A^2) \frac{I}{e \Gamma_{\rm ph}}
        + \lambda_A^2 \langle n_e \rangle.
        \label{eq:N_A}
\end{align}
The first two terms in Eq.\ (\ref{eq:N_A})
indicate the emission of $\nu$ phonons by the
phonon-assisted tunneling (from dot $L$ to dot $R$)
and creation of $2 \lambda_A^2$ phonons by lattice distortion
(with two tunnelings between the DQD and leads)
per transfer of a single electron  through the DQD.
The last term describes the average number of polarons
$\langle n_e \rangle$ in the stationary state.

When $\Gamma_{L,R} \gg \Gamma_{\rm ph}$, we obtain
$I = I_0 + \mathcal{O}(\Gamma_{\rm ph}/\Gamma_{L,R})$,
where $I_0 = e \Gamma_R / (2 + \alpha)$
is the current at the main peak in the absence of
electron-phonon interaction, and
    \cite{supplement}
\begin{align}
    g^{(2)}_A (0) = \frac{\nu + 4 \lambda_A^2}{\nu + 2 \lambda_A^2}
        + \mathcal{O} (\Gamma_{\rm ph}/\Gamma_{L,R}).
        \label{eq:g2_A}
\end{align}
These explain the numerical results in Fig.\ \ref{fig:Delta}
at the current subpeaks.
Equation (\ref{eq:g2_A})
indicates $g^{(2)}_A (0) \simeq 1$ (phonon lasing)
for $\lambda_A^2 \ll \nu$ and
$g^{(2)}_A (0) \simeq 2$ (phonons thermalized by lattice
distortion) for $\lambda_A^2 \gg \nu$. In the latter case,
the phonons follow the Bose distribution with $T^*$
for the deduction of $\langle N_A \rangle$ in Eq.\ (\ref{eq:N_A}).

\begin{figure}[t]
\begin{center}
    \includegraphics[width=8cm]{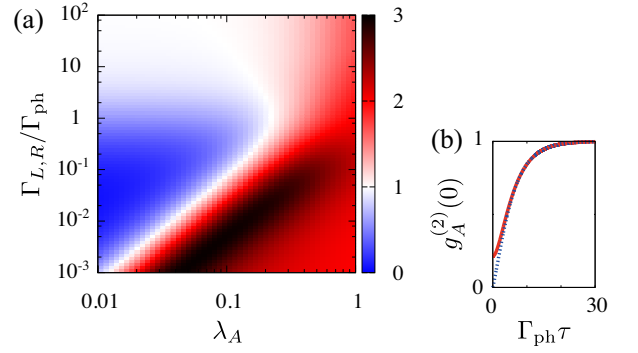}
    \caption{(Color online).
    (a) Color-scale plot of autocorrelation function of $A$-phonons,
    $g^{(2)}_A (0)$, at the first subpeak of current in a plane
    of electron-phonon coupling $\lambda_A$ and
    $\Gamma_{L, R}/\Gamma_{\rm ph}$.
    $\Gamma_L = \Gamma_R \equiv \Gamma_{L, R}$,
    $\lambda_S = 0$, and $V_C = 0.1~\hbar \omega_{\rm ph}$.
    (b) $g^{(2)}_A (\tau)$ at $\lambda_A = 0.05$ and
    $\Gamma_{L,R} = 0.1~\Gamma_{\rm ph}$ as a function
    of $\tau$ (solid line). The electric current autocorrelation
    function $g^{(2)}_{\rm current} (\tau)$ is also shown by a
    dotted line.
    \label{fig:antibunching}}
\end{center}
\end{figure}

Thus far, we have discussed the phonon lasing in the
case of $\Gamma_{L, R} \gg \Gamma_{\rm ph}$.
If the tunnel coupling
is tuned to be $\Gamma_{L, R} \lesssim \Gamma_{\rm ph}$,
we observe another phenomenon, i.e., antibunching of
LO phonons \cite{Lambert2008}.
Figure \ref{fig:antibunching}(a) presents a color-scale plot of
$g^{(2)}_A(0)$ at the first current subpeak 
in the $\lambda_A$--$(\Gamma_{L, R}/\Gamma_{\rm ph})$ plane for
$\Gamma_L = \Gamma_R \equiv \Gamma_{L, R}$ and $\lambda_S=0$.
At $\lambda_A=0.05$ and $\Gamma_{L, R}/\Gamma_{\rm ph}=0.1$, for example,
$g^{(2)}_A (0) \ll 1$, representing a strong antibunching of phonons.
This is because the phonon emission is regularized by the
electron transport through the DQD.
In Fig.\ \ref{fig:antibunching}(b),
we plot the autocorrelation function of the electric current
\begin{align}
    g^{(2)}_{\rm current} (\tau) =
    \langle :n_R(0) n_R(\tau): \rangle/ \langle n_R \rangle^2,
\end{align}
where $n_R$ is the electron number in dot $R$. It fulfills 
$g^{(2)}_{\rm current} (0)=0$, indicating the antibunching of
electron transport, since dot $R$ is empty just after
the electron tunnels out \cite{Emary2012}.
Remarkably, $g^{(2)}_A(\tau)$ almost coincides
with $g^{(2)}_{\rm current} (\tau)$.
Again at very strong couplings $\lambda_A \gtrsim 1$, neither
phonon antibunching nor phonon lasing can be
observed  because of an effective phonon thermalization
due to the Franck-Condon effect.

In our calculations, we have neglected the $S$-phonon coupling
that, however, does not affect
the dynamics of $A$-phonons
as we have checked.
We have also disregarded acoustic-phonon-assisted tunneling to
excited levels in dot $R$ because the energy of LA phonons at
small wave numbers
$|{\bm q}| \lesssim 1/\mathcal{R}$ is comparable to or
lower than the spacing between the energy levels for $
\mathcal{R}<$ 100 nm,
and the coupling to LA phonons is much weaker than that to LO phonons.
Indeed, the LO-phonon-assisted transport was clearly observed
for level spacings $\Delta$ tuned to $\hbar \omega_{\rm ph}$ and
$2\hbar \omega_{\rm ph}$ in recent experiments
    \cite{Amaha2012}.

Finally, we discuss possible experimental realizations to
observe LO phonon lasing and antibunching in semiconductor-based DQDs.
In GaAs, an LO phonon around the $\Gamma$ point decays into
an LO phonon and a TA phonon around the L point,
which are not coupled to the DQD.
These daughter phonons can be detected by the transport through another
DQD fabricated nearby
    \cite{Gasser2009,Harbusch2010}.
Alternatively, the modulation of the dielectric constant by the
phonons could be observed by near-field spectroscopy
    \cite{Cunningham2008}.
With a decay rate $\Gamma_{\rm ph} \sim 0.1~\text{THz}$ in GaAs
    \cite{Vallee1994},
however, 
the lasing condition $\Gamma_{L, R} \gg \Gamma_{\rm ph}$
might be difficult to realize. Other materials with a longer lifetime of
optical phonons, such as ZnO \cite{Aku-Leh2005}, may be preferable for
observing the phonon lasing.

Our fundamental research of LO phonon statistics is also applicable
to a freestanding semiconductor membrane as a phonon cavity
\cite{Weig2004,Ogi2010},
in which a resonating mode plays a role of LO phonons. Our theory
implies that a DQD could generate various quantum states of
mechanical oscillators.

\acknowledgements

The authors acknowledge fruitful discussion with
K.\ Ono, S.\ Amaha,
Y.\ Kayanuma, K.\ Saito,
C.\ P\"oltl, T.\ Yokoyama, and A.\ Yamada.
This work was partially supported by KAKENHI (No.\ 23104724 and No.\ 24-6574),
the Institutional Program for Young Researcher Oversea Visits
and the International Training Program
from the Japan Society for the Promotion of Science,
the Graduate School Doctoral Student Aid Program from Keio University,
and the German DFG via SFB 910 and project BR 1528/8-1.

\begin{widetext}

\newpage

\begin{center}
    \large \bf
    Supplementary Material for\\
    ``Optical Phonon Lasing in Semiconductor Double Quantum Dots''
\end{center}

\begin{center}
    Rin Okuyama$^1$, Mikio Eto$^1$, and Tobias Brandes$^2$ \\
\end{center}

\begin{center}
    {\em $^1$Faculty of Science and Technology,
        Keio University, Yokohama 223-8522, Japan} \\
    {\em $^2$Institut f\"ur Theoretische Physik,
        Technische Universit\"at Berlin, D-10623 Berlin, Germany} \\
\end{center}

\vspace{5mm}

In this supplemental material,
we derive analytical expressions for the current $I$,
number of phonons $\langle N_A \rangle$, and
autocorrelation function of phonons $g^{(2)}_A (0)$ at the current subpeaks
in Eqs.\ (15)--(17) in the main material.
When $\Delta \simeq \nu \hbar \omega_{\rm ph}$,
the energy eigenstates are given by the zero-electron states
$
    |0, n \rangle = |0 \rangle \otimes |n \rangle_A,
$
bonding and anti-bonding states between polarons
$
    |\pm, n \rangle = \frac{1}{\sqrt{2}} \left( |L, n \rangle
        \pm |R, n + \nu \rangle \right),
$
and polarons localized in dot $R$,
$
    |R, n \rangle
$
$
    (n = 0, 1, 2, \ldots, \nu - 1),
$
in a good approximation for $V_C \ll \hbar \omega_{\rm ph}$,
as mentioned in the main material.
We have introduced polaron states
$
    |L (R), n \rangle = |L (R) \rangle \otimes
        \mathcal{T}^{(\dagger)} |n \rangle_A,
$
with
$
    \mathcal{T} = e^{- \lambda_A (a_A^\dagger - a_A)}
$.
The density matrix is given by
\begin{align}
    \rho = \sum_{n = 0}^\infty P_{0, n} |0, n \rangle \langle 0, n|
        + \sum_{\sigma = \pm} \sum_{n = 0}^\infty
            P_{\sigma, n} |\sigma, n \rangle \langle \sigma, n|
        + \sum_{n = 0}^{\nu - 1} P_{R, n} |R, n \rangle \langle R, n|
\end{align}
in the Born-Markov-Secular approximation.
We define occupation number operators for zero-electron states,
bonding and anti-bonding states between the polarons,
and polarons localized in dot $R$ as
\begin{align}
    n_0 = \sum_{n = 0}^\infty |0, n \rangle \langle 0, n|
        = |0 \rangle \langle 0|,
        \qquad
    n_{\rm mol} = \sum_{\sigma = \pm} \sum_{n = 0}^\infty
        |\sigma, n \rangle \langle \sigma, n|,
        \qquad
    \tilde n_R = \sum_{n = 0}^{\nu - 1} |R, n \rangle \langle R, n|,
\end{align}
respectively. The relation of
$n_0 + n_{\rm mol} + \tilde n_R=1$ holds.
The electron number in the DQD is given by
$n_e = n_{\rm mol} + \tilde n_R = 1 - n_0$.
The expectation values of the occupation numbers are
\begin{align}
    \langle n_0 \rangle = \sum_{n = 0}^\infty P_{0, n},
        \qquad
    \langle n_{\rm mol} \rangle = \sum_{n = 0}^\infty P_{{\rm mol}, n},
        \qquad
    \langle \tilde n_R \rangle = \sum_{n = 0}^{\nu - 1} P_{R, n}.
\end{align}

In the stationary state,
the rate equations in Eqs.\ (12)--(14) in the main material yield
\begin{align}
    0 &= -\Gamma_L P_{0, n}
        + \sum_{m = 0}^{\infty} \frac{\Gamma_R}{2}
        |_A \langle n | \mathcal{T}^\dagger | m + \nu \rangle_A|^2
        P_{\text{mol}, m} + \sum_{m = 0}^{\nu - 1} \Gamma_R
        |_A \langle n | \mathcal{T}^\dagger | m \rangle_A|^2 P_{R, m}
           + \Gamma_{\rm ph} \bigl[ (n + 1) P_{0, n + 1} - n P_{0, n} \bigr],
        \label{eq:app_rate_zero} \\
    0 &= - \frac{\Gamma_R}{2} P_{\text{mol}, n}
        + \sum_{m = 0}^{\infty} \Gamma_L
        |_A \langle n | \mathcal{T} | m \rangle_A|^2 P_{0, m}
        + \Gamma_{\rm ph} \left [
        \left( n + 1 + \frac{\nu}{2} \right) P_{\text{mol}, n + 1}
        - \left( n + \frac{\nu}{2} \right)
        P_{\text{mol}, n} \right],
        \label{eq:app_rate_mol}
\end{align}
with $P_{{\rm mol}, n} = P_{+, n} + P_{-, n}$ $(n = 0, 1, 2, \ldots)$, and
\begin{align}
    0 &= - \Gamma_R P_{R, n}
        + \Gamma_{\rm ph} \bigl[ (n + 1) P_{R, n + 1} - n P_{R, n} \bigr],
        \label{eq:app_rate_R}
\end{align}
with $P_{R, \nu} = P_{{\rm mol}, 0} / 2$
$(n = 0, 1, 2, \ldots, \nu - 1)$.

\section{Current and Electron Number}

First, we express the current
$I = e \Gamma_L \langle n_0 \rangle$
in the stationary state.
For the purpose,
we sum up both sides of Eq.\ (\ref{eq:app_rate_zero}) over $n$.
Using
\begin{align}
    \sum_{n = 0}^\infty \left|_A
        \langle n | \mathcal{T}^\dagger
        | m \rangle_A \right|^2
    &= \left._A \langle m | \mathcal{T}
        \left( \sum_n |n \rangle_{AA} \langle n| \right)
        \mathcal{T}^\dagger | m \rangle_A \right.
    = 1,
\end{align}
we obtain
\begin{align}
    0 = -\Gamma_L \langle n_0 \rangle
        + \frac{\Gamma_R}{2} \langle n_{\rm mol} \rangle
        + \Gamma_R \langle \tilde n_R \rangle.
\end{align}
With 
$\langle n_0 \rangle + \langle n_{\rm mol} \rangle
+ \langle \tilde n_R \rangle=1$,
we derive
\begin{align}
    \langle n_0 \rangle = \frac{\alpha}{2 + \alpha}
        (1 + \langle \tilde n_R \rangle),
        \qquad
    \langle n_{\rm mol} \rangle = \frac{2}{2 + \alpha}
        \left[ 1 - (1 + \alpha) \langle \tilde n_R \rangle \right],
\end{align}
with $\alpha = \Gamma_R / \Gamma_L$.
These equations result in Eq.\ (15) in the main material:
\begin{align}
    I = e \Gamma_R \frac{1 + \langle \tilde n_R \rangle}{2 + \alpha}, \qquad
    \langle n_e \rangle = \frac{2 - \alpha \langle \tilde n_R \rangle}
        {2 + \alpha}.
\end{align}
The summation of Eq.\ (\ref{eq:app_rate_R}) over $n$ yields
\begin{align}
    \langle \tilde n_R \rangle = \frac{\nu \Gamma_{\rm ph}}{2 \Gamma_R}
        P_{{\rm mol}, 0}.
        \label{eq:n_R}
\end{align}

\section{Phonon Number}

Next, we derive the phonon number which is
\begin{align}
    \langle N_A \rangle
    &= \sum_{n = 0}^\infty P_{0, n} \langle 0, n| N_A |0, n \rangle
        + \sum_{\sigma = \pm} \sum_{n = 0}^\infty P_{\sigma, n}
            \langle \sigma, n | N_A |\sigma, n \rangle
        + \sum_{n = 0}^{\nu - 1} P_{R, \nu} \langle R, n| N_A |R, n \rangle \\
    &= \sum_{n = 0}^\infty n P_{0, n} + \sum_{n = 0}^\infty
        \left( n + \frac{\nu}{2} + \lambda_A^2 \right) P_{{\rm mol}, n}
        + \sum_{n = 0}^{\nu - 1} (n + \lambda_A^2) P_{R, n} \\
    &\equiv \langle N_A n_0 \rangle
        + \langle N_A n_{\rm mol} \rangle + \langle N_A \tilde n_R \rangle.
\end{align}
We have used
\begin{align}
    \mathcal{T^\dagger} N_A \mathcal{T}
        = ( \mathcal{T^\dagger} a_A^\dagger \mathcal{T} )
            ( \mathcal{T^\dagger} a_A \mathcal{T} )
        = (a_A^\dagger - \lambda_A) (a_A - \lambda_A).
\end{align}
We multiply both sides of Eqs.\ (\ref{eq:app_rate_zero})--(\ref{eq:app_rate_R})
by $n$ and sum up over $n$:
\begin{align}
    0 &= - (\Gamma_L + \Gamma_{\rm ph}) \langle N_A n_0 \rangle
        + \frac{\Gamma_R}{2} \langle N_A n_{\rm mol} \rangle
        + \Gamma_R \langle N_A \tilde n_R \rangle
        + \frac{\nu \Gamma_R}{4} \langle n_{\rm mol} \rangle,
        \label{eq:N_A_zero} \\
    0 &= \Gamma_L \langle N_A n_0 \rangle
        - \left( \frac{\Gamma_R}{2} +\Gamma_{\rm ph} \right)
            \langle N_A n_{\rm mol} \rangle
        + \lambda_A^2 \Gamma_L \langle n_0 \rangle
        + \left( \frac{\nu + 2 \lambda_A^2}{4} \Gamma_R
            + \lambda_A^2 \Gamma_{\rm ph} \right) \langle n_{\rm mol} \rangle
        + \Gamma_R \langle \tilde n_R \rangle,
        \label{eq:N_A_mol} \\
    0 &= - (\Gamma_R + \Gamma_{\rm ph}) \langle N_A \tilde n_R \rangle
    + \left[ (\nu - 1 + \lambda_A^2) \Gamma_R
        + \lambda_A^2 \Gamma_{\rm ph} \right] \langle \tilde n_R \rangle.
        \label{eq:N_A_R}
\end{align}
Here, we have used
\begin{align}
    \sum_{n = 0}^\infty n \left|_A
        \langle n | \mathcal{T}^\dagger
        | m \rangle_A \right|^2
    = \left._A \langle m | \mathcal{T}
        N_A \left( \sum_n |n \rangle_{AA} \langle n| \right)
        \mathcal{T}^\dagger | m \rangle_A \right.
    = \sum_{m} \left._A \langle m | \mathcal{T} N_A \mathcal{T}^\dagger
    | m \rangle_A \right..
\end{align}
From Eqs.\ (\ref{eq:N_A_zero})--(\ref{eq:N_A_R}), we obtain
Eq.\ (16) in the main material:
\begin{align}
    \langle N_A \rangle = (\nu + 2 \lambda_A^2) \frac{I}{e \Gamma_{\rm ph}}
        + \lambda_A^2 \langle n_e \rangle.
\end{align}

\section{Phonon Autocorrelation Function}

Finally, we derive the phonon autocorrelation function at zero time delay,
\begin{align}
    g^{(2)}_A (0) = \frac{\langle :N_A^2: \rangle}{\langle N_A \rangle^2}
        = \frac{\langle N_A^2 \rangle - \langle N_A \rangle}
            {\langle N_A \rangle^2},
\end{align}
where
\begin{align}
    \langle N_A^2 \rangle
    &= \sum_{n = 0}^\infty P_{0, n} \langle 0, n| N_A^2 |0, n \rangle
        + \sum_{\sigma = \pm} \sum_{n = 0}^\infty P_{\sigma, n}
            \langle \sigma, n | N_A^2 |\sigma, n \rangle
        + \sum_{n = 0}^{\nu - 1} P_{R, \nu} \langle R, n| N_A^2 |R, n \rangle \\
    &= \sum_{n = 0}^\infty n^2 P_{0, n} + \sum_{n = 0}^\infty
        \left[ n^2 + \lambda_A^2 (4n + 1 + \lambda_A^2)
        + \nu \left( n + \frac{\nu}{2} + 2 \lambda_A^2 \right) \right]
        P_{{\rm mol}, n}
     + \sum_{n = 0}^{\nu - 1}
        \left[ n^2 + \lambda_A^2 (4 n + 1 + \lambda_A^2) \right]
        P_{R, n} \\
    &\equiv \langle N_A^2 n_0 \rangle
        + \langle N_A^2 n_{\rm mol} \rangle + \langle N_A^2 \tilde n_R \rangle.
\end{align}
We multiply both sides of Eqs.\ (\ref{eq:app_rate_zero})--(\ref{eq:app_rate_R})
by $n^2$ and sum up over $n$.
A similar technique to the last section leads to
\begin{align}
    0 &= -(\Gamma_L + 2 \Gamma_{\rm ph}) \langle N_A^2 n_0 \rangle
        + \frac{\Gamma_R}{2} \langle N_A^2 n_{\rm mol} \rangle
        + \Gamma_R \langle N_A^2 \tilde n_R \rangle
        + \Gamma_{\rm ph} \langle N_A n_0 \rangle
        + \frac{\nu \Gamma_R}{2} \langle N_A n_{\rm mol} \rangle
        + \frac{\nu \lambda_A^2 \Gamma_R}{2} \langle n_{\rm mol} \rangle,
        \label{eq:N_A_2_zero} \\
    0 &= \Gamma_L \langle N_A^2 n_0 \rangle
        - \left( \frac{\Gamma_R}{2} + 2 \Gamma_{\rm ph} \right)
        \langle N_A^2 n_{\rm mol} \rangle
        + 4 \lambda_A^2 \Gamma_L \langle N_A n_0 \rangle
        + \left[ \frac{\nu + 4 \lambda_A^2}{2} \Gamma_R
            + ( \nu + 1 + 8 \lambda_A^2) \Gamma_{\rm ph} \right]
            \langle N_A n_{\rm mol} \rangle
        \nonumber \\
    &\quad + \lambda_A^2 (1 + \lambda_A^2) \Gamma_L \langle n_0 \rangle
        + \lambda_A^2 \left[ \frac{1 - \nu - 3 \lambda_A^2}{2} \Gamma_R
            + (1 - \nu - 6 \lambda_A^2) \Gamma_{\rm ph} \right]
            \langle n_{\rm mol} \rangle
        - \Gamma_R \langle \tilde n_R \rangle,
        \label{eq:N_A_2_mol} \\
    0 &= -(\Gamma_R + 2 \Gamma_{\rm ph}) \langle N_A^2 \tilde n_R \rangle
        + \left[ 4 \lambda_A^2 \Gamma_R
            + (1 + 8 \lambda_A^2) \Gamma_{\rm ph} \right]
            \langle N_A \tilde n_R \rangle
        \nonumber \\
    &\quad + \left\{ \left[ (\nu - 1)^2 + \lambda_A^2 (1 - 3\lambda_A^2) \right]
                \Gamma_R
            + \lambda_A^2 ( 1 - 6 \lambda_A^2) \Gamma_{\rm ph} \right\}
            \langle \tilde n_R \rangle
        \label{eq:N_A_2_R}.
\end{align}
From Eqs.\ (\ref{eq:N_A_2_zero})--(\ref{eq:N_A_2_R}), we find
\begin{align}
    \langle N_A^2 \rangle - \langle N_A \rangle
    &= 2 \lambda_A^2 \frac{\Gamma_L}{\Gamma_{\rm ph}} \langle N_A n_0 \rangle
        + \left( \frac{\nu + 2 \lambda_A^2}{2} \frac{\Gamma_R}{\Gamma_{\rm ph}}
            + \frac{\nu + 4 \lambda_A^2}{2} \right)
            \langle N_A n_{\rm mol} \rangle
        + 2 \lambda_A^2 \left( \frac{\Gamma_R}{\Gamma_{\rm ph}} + 2 \right)
            \langle N_A \tilde n_R \rangle
        \nonumber \\
    & \quad - \frac{\nu + 2 \lambda_A^4}{2} \frac{\Gamma_L}{\Gamma_{\rm ph}}
            \langle n_0 \rangle
        - \frac{\lambda_A^2 (\nu + 6 \lambda_A^2)}{2}
            \langle n_{\rm mol} \rangle
        - \left[ \frac{\nu (2 - \nu)}{2} \frac{\Gamma_R}{\Gamma_{\rm ph}}
            + 3 \lambda_A^2 \right] \langle \tilde n_R \rangle.
\end{align}

Now we evaluate $g^{(2)}_A (0)$ in the case of
$\Gamma_{L, R} \gg \Gamma_{\rm ph}$. In this case,
$
    \langle \tilde n_R \rangle = \mathcal{O} (\Gamma_{\rm ph} / \Gamma_{L, R})
$
from Eq.\ (\ref{eq:n_R}). Then
\begin{align}
    I = \frac{e \Gamma_R}{2 + \alpha}
        + \mathcal{O}(\Gamma_{\rm ph} / \Gamma_{L, R})
\end{align}
and
\begin{align}
    \langle N_A \rangle
    = \frac{\nu + 2 \lambda_A^2}{2 + \alpha}
                \frac{\Gamma_R}{\Gamma_{\rm ph}}
        + \mathcal{O}(1).
\end{align}
Equations (\ref{eq:N_A_zero}) and (\ref{eq:N_A_R}) yield
\begin{align}
    2 \langle N_A n_0 \rangle
        = \alpha \langle N_A n_{\rm mol} \rangle
            + \mathcal{O} (1), \qquad
    \langle N_A \tilde n_R \rangle
        = \mathcal{O}(\Gamma_{\rm ph} / \Gamma_{L, R}).
\end{align}
Using
$
    \langle N_A \rangle
    = \langle N_A n_0 \rangle
        + \langle N_A n_{\rm mol} \rangle
        + \langle N_A \tilde n_R \rangle
$, we have
\begin{align}
    \langle N_A n_0 \rangle
        = (\nu + 2 \lambda_A^2) \frac{\alpha}{(2 + \alpha)^2}
            \frac{\Gamma_R}{\Gamma_{\rm ph}} + \mathcal{O} (1),
        \qquad
    \langle N_A n_{\rm mol} \rangle
        = (\nu + 2 \lambda_A^2) \frac{2}{(2 + \alpha)^2}
                    \frac{\Gamma_R}{\Gamma_{\rm ph}} + \mathcal{O}(1).
\end{align}
Using these relations, we obtain Eq.\ (17) in the main material:
\begin{align}
    g^{(2)}_A (0) = \frac{\nu + 4 \lambda_A^2}{\nu + 2 \lambda_A}
        + \mathcal{O} (\Gamma_{\rm ph} / \Gamma_{L, R}).
\end{align}

\end{widetext}


\begin{thebibliography}{10}
\bibitem{com:flat_dispersion}
If a weak quadratic dispersion around the $\Gamma$ point is taken
into account, the collective phonon modes are scattered by the rate of
$\gamma \sim \partial^2 \omega_{\rm ph} / \partial q^2 |_{q = 0}
/ \mathcal{R}^2$,
which is lower than the decay rate $\Gamma_{\rm ph}$ of
LO phonons by two orders of magnitude in GaAs quantum dots
with $\mathcal{R} = 10 \sim 100~\text{nm}$.
\bibitem{com:large_bias}
When the bias voltage is not sufficiently large, the current is suppressed
by the Franck-Condon blockade.
Here, we assume a large bias to avoid the blockade.
\end{thebibliography}
\end{document}